\documentclass[aps,preprint,a4paper,showkeys,showpacs]{revtex4-1}
\usepackage{amsmath}
\usepackage{graphicx}
\usepackage{natbib}
\usepackage{color}
\usepackage{textcase}
\usepackage{url}
\usepackage{bm}

\begin{document}
\title{Boundary-Induced Pattern Formation from Temporal Oscillation: Spatial Map Analysis}

\author{Takahiro Kohsokabe}
\affiliation{Department of Basic Science, Graduate School of Arts and Sciences, The University of Tokyo, 3-8-1 Komaba, Meguro, Tokyo 153-8902, Japan}
\author{Kunihiko Kaneko}
\affiliation{Research Center for Complex Systems Biology, Graduate School of Arts and Sciences, The University of Tokyo, 3-8-1 Komaba, Meguro, Tokyo 153-8902, Japan}
\begin{abstract}
Boundary-induced pattern formation from a spatially uniform state is investigated using one-dimensional reaction-diffusion equations. The temporal oscillation is successively transformed into a spatially periodic pattern, triggered by diffusion from the fixed boundary. 
We introduced a spatial map, whose temporal sequence, under selection criteria from multiple stationary solutions, can completely reproduce the emergent pattern, by replacing the time with space.
The relationship of the pattern wavelength with the period of oscillation is also obtained. The generality of the pattern selection process and algorithm is discussed with possible relevance to biological morphogenesis.
\end{abstract}
\maketitle

Spatial patterns are ubiquitous in non-equilibrium systems, and the process of spontaneous pattern formation has been extensively studied as one of the main issues in non-linear dynamics. Topic areas in pattern formation include fluid, solid-state, optical, geophysical, chemical, and biological systems\cite{nicolis1977self,cross1993pattern,walgraef1997spatio,winfree2001geometry,kapral2012chemical}.
\par In particular, the reaction-diffusion system, which was pioneered by Turing\cite{turing1952chemical}, has been a focus of non-equilibrium studies. In his celebrated study, Turing showed how a spatial pattern or temporal rhythm is spontaneously formed from a spatially homogeneous and temporally stationary state, which is unstable against perturbations. Turing classified such pattern formation processes into six cases, one of which is known as the \textit{Turing pattern}; stationary periodic pattern with the finite wavelength given by linear stability analysis.
 While Turing originally proposed his theory as a model of morphogenesis, it has been applied to general pattern formation dynamics beyond developmental biology\cite{segel1972dissipative, LENGYEL650}.
\par Spontaneous pattern formation from temporally dynamic states is also possible in a reaction-diffusion system. Destabilization of a spatially homogeneous and temporally oscillatory state by diffusive interaction often results in spatiotemporal dynamics such as waves, spirals, and turbulence\cite{kuramoto1984chemical}. Spatiotemporal dynamics by combination of Turing instability for pattern formation and Hopf bifurcation for temporal oscillation have been studied as Turing-Hopf bifurcation\cite{PhysRevE.54.261,meixner1997generic,baurmann2007instabilities}.
\par However, an important factor that has not been fully explored is the influence of boundary conditions on spatial patterns. In particular, pattern formation from temporally dynamic states may be crucially influenced by the introduction of a fixed boundary. The influence may be globally propagated to alter the pattern dynamics. (See also \cite{pomeau1981wavelength, fujimoto2001sensitive} for the relevance of boundary conditions in pattern dynamics.)
In the present letter, we are concerned with such boundary-induced pattern formation, namely, how a fixed boundary condition destabilizes the stable, temporally periodic, spatially uniform state and transforms it into a temporally stationary, spatially periodic pattern.
\par The question of boundary-induced pattern formation is not only of theoretical interest but also of experimental interest in developmental biology.
 In the somitogenesis of vertebrate development, temporal oscillation in protein expression is fixed into a spatial pattern\cite{palmeirim1998uncoupling,pourquie2001vertebrate}. As development progresses, the intracellular concentration of the protein c-hairy1, which initially oscillates in the system, is fixed into a striped pattern along aligned cells. So far, such spatial pattern formation has been studied by introducing external inputs that move in space with time development (\textit{clock and wavefront})\cite{cooke1976clock}. Diffusive interaction has been introduced by Meinhardt in addition to external input using a spatial gradient\cite{meinhardt1982models}.
  For these theoretical studies, the use of an external input is essential to destabilize the homogeneous state for pattern formation. Some recent experiments, however, suggested that an external input may not be essential, and intrinsic instability due to diffusion might result in somitogenesis\cite{dias2014somites,cotterell2015local}. If boundary-induced pattern formation is possible without imposing an external input throughout the space, it will provide a plausible mechanism for vertebrate somitogenesis.
\par Here, we demonstrate that temporal oscillation in a one-dimensional reaction-diffusion system is fixed into a stationary periodic pattern by introducing fixed boundary condition. In contrast to the celebrated Turing pattern, the wavelength of the generated pattern cannot be obtained using linear stability analysis around the fixed point. Instead, to predict the selected pattern, we introduced a one-dimensional spatial map, whose attractor gives the one-dimensional pattern by replacing time with space. The generality of this oscillation fixation and the pattern selection mechanism will be discussed.

Now, consider a one-dimensional reaction-diffusion system of two components $X$ and $Y$. We assume that the diffusion of $X$ is much faster than that of $Y$, and the diffusion of the latter is neglected for simplicity (the formalism to be discussed is valid even without this approximation). Then, the equation is written as
\begin{equation}
\left.
\begin{array}{l}
\dfrac{\partial X}{\partial t}=f(X,Y)+D\dfrac{{\partial}^2 X}{\partial x^2}\\[8pt]\
\dfrac{dY}{dt}=g(X,Y)\\[12pt]
\end{array}
\right.
\label{eq:reaction-diffusion}
\end{equation}
where $f(X,Y)$ and $g(X,Y)$ are the reaction functions for $X$ and $Y$, $D$ is the diffusion constant of $X$, and the attractor of the dynamical system without $D$ is a limit cycle\footnote{When this limit cycle is generated by a Hopf bifurcation from the fixed point $(X^*,Y^*)$, the eigenvalues of the Jacobi matrix around the fixed point satisfy the positive real part with a non-zero imaginary part so that  $a+d>0$ and $-2<\frac{d-a}{\sqrt{-bc}}<2$\cite{turing1952chemical}, where 
\[
\begin{pmatrix}
a & b \\ c &d
\end{pmatrix}
=
\begin{pmatrix}
\frac{\partial f}{\partial X} & \frac{\partial f}{\partial Y}\\
\frac{\partial g}{\partial X} & \frac{\partial g}{\partial Y}
\end{pmatrix}_{X=X^*, Y=Y^*}
\]}
\label{eq:Jacobian}.
\par
We consider the case in which the spatially uniform, limit cycle state is stable against perturbations so that, under Neumann or periodic boundary conditions, this state is the attractor.
However, under a fixed boundary condition, the variable $X$ close to the boundary cannot oscillate, which may destabilize the oscillatory attractor. Indeed, we have found that a fixed periodic pattern is often generated in this case.

As a specific example, consider the following system, 
\begin{equation}
\left.
\begin{array}{l}
f(X,Y)=\dfrac{1}{1+e^{-\beta (Y-1/2)}}-X\\[8pt]
g(X,Y)=\dfrac{1}{1+e^{-\beta (Y-X)}}-Y\\
\end{array}
\right.
\label{eq:example1}
\end{equation}
which describe the protein expression dynamics with two genes, where $X$ inhibits the expression of $Y$ and $Y$ activates the expressions of both $X$ and $Y$\cite{mjolsness1991connectionist,PhysRevE.88.032718}.
Without the diffusion term, this system has one unstable fixed point $(X^*,Y^*)=(1/2,1/2)$ if $\beta>8$, and the limit cycle is an attractor. Indeed, from initial conditions close to a spatially homogenous state, a uniform, limit-cycle state is reached if the boundary condition is Neumann or periodic. In contrast, if a fixed boundary condition is adopted for at least one end, i.e., $X(0)=X_0$, temporal oscillation is replaced by a fixed, spatially periodic pattern \footnote{A boundary condition for $Y$ is not required as diffusion in $Y$ is neglected.} (see Fig. 1A and 1B where $\frac{\partial X}{\partial x}|_{x=L}=0$ is adopted at the other end). Note that a pattern of the same wavelength is organized independently of $X_0$ as long as $X_0$ is between [0,1](see Fig. 3).
Here, oscillation ceased in the vicinity of $x=0$, which works as a boundary for $(X(x),Y(x))$ for slightly larger $x$. The oscillation is successively fixed for larger $x$. 

We analytically examined how the organized pattern is determined. First, we confirmed that the linear stability analysis around the uniform fixed-point solution cannot explain the wavelength in contrast to the Turing instability case. The real part of eigenvalue is positive against perturbations of a given wavenumber $k$ around the unstable fixed point solution $(0.5,0.5)$ for $k<k_1=\frac{1}{4\pi}\sqrt{\frac{\beta-8}{D}}$, while the imaginary part is nonzero. On the other hand, for $k>k_2=\frac{1}{4\pi}\sqrt{\frac{\beta^2 -4\beta+16}{D(\beta-4)}}$, one of the eigenvalues is positive with a vanishing imaginary part where Turing instability exists.
 Thus, both Hopf and Turing instability coexist in this system (see Supplemental Figure 1A).
Then, we contrasted the marginal values $k_1$ and $k_2$ with the wavenumber of the emergent pattern by changing $\beta$. The observed wavelength agrees neither with $k_1$ nor with $k_2$(see Supplemental Figure 1B). Even the dependence of the parameters on $\beta$ does not agree. Indeed, this discrepancy is natural: pattern formation occurs from the homogeneous limit cycle, which is far from the unstable fixed solution. Hence, the standard analysis for the Turing pattern does not work in this system.\par

Now we need to find a procedure to determine the spatial pattern without utilizing linear stability analysis. We note that once the diffusion term is given, the stationary solution $(X_{st}(x), Y_{st}(x))$ can be obtained from the dynamical system of two variables $X,Y$, where standard nullcline analysis works. As shown in Fig. 2B, only the nullcline of $X$ (not that of $Y$) is horizontally shifted with the diffusion term. Accordingly, the fixed point ($X_{st}(x),Y_{st}(x))$ is shifted in the state space.

 Once the fixed point $(X_{st}(x), Y_{st}(x))$ is given, the gradient term is obtained. For a fixed pattern, the fixed point $(X_{st}(x),Y_{st}(x))$ and the gradient term are determined self-consistently.

This self-consistent condition, given by $\partial X(x, t)/\partial t=0, d Y(x,t)/d t = 0$, can be solved explicitly by adopting spatial discretization to compute the diffusion term. In general, this condition is given by
\begin{equation}
\left.
\begin{array}{l}
f(X_l,Y_l)+D^{'}(X_{l+1}-2 X_l+X_{l-1})=0\\
g(X_l,Y_l)=0\\[10pt]
\end{array}
\right.
\label{eq:discretized_equation}
\end{equation}
where $l$ is discretized space index with $a$ as the discretized unit length, i.e., $x=al$ and $X_l=X(x/a)$, $Y_l=Y(x/a)$. Here, $D^{'}$ is the rescaled diffusion constant which satisfies $D{'}={D}/{a^2}$. From these equations, we can derive the spatial map as follows:
\begin{equation}
\left.
\begin{array}{l}
X_{l+1}=2X_l-X_{l-1}-\dfrac{f(X_l, Y_l)}{D^{'}}\\
Y_l=g^{-1}_{X_l}(0)\\
\end{array}
\right.
\label{eq:spatialmap}
\end{equation}
where $g^{-1}_{X_l}$ is the inverse function of $g(X,Y)$ given that $X$ is fixed to $X_l$, which corresponds to the cross-points of the nullclines. With these equations, $X_{l+1}$ is determined from $X_l$, $X_{l-1}$, and $Y_{l}$, while $Y_{l+1}$ is determined as $g^{-1}_{X_{l+1}}(0)$ (for the use of a ``spatial map'' in the analysis of a spatial pattern, see also \cite{aubry1978new,willeboordse1995pattern}). However, there can be multiple candidate solutions for $g^{-1}_{X_{l+1}}(0)$. In fact, not all of the above solutions are stable in terms of dynamical systems (\ref{eq:reaction-diffusion}). Furthermore, this solution must be attracted from the uniform oscillatory state. These properties lead to the following selection principles.\\
\\\textbf{Selection principles}
\begin{enumerate}
\item Ignore candidates in which the resulting state $(X_{l+1},Y_{l+1})$ is not stable against small perturbations in $X_{l+1}$.
\item Ignore candidates in which the resulting value $X_{l+2}$ is out of the basin of the original limit cycle.
\item Calculate $X_{l+2}$ from all remaining candidates, and choose the solution that gives the smallest value of $|X_{l+2}-2X_{l+1}+X_l|$.
\end{enumerate}
The first criterion is necessary for the stability of the novel fixed point. The second criterion is necessary for the pattern to emerge from the limit cycle. The third criterion implies the smallest net diffusion, which minimizes the distance from a uniform state. With the second and third criteria, the solution that is attracted from the uniform oscillatory state is selected.

An example of the pattern obtained from the spatial map with the above selection criteria is shown in Fig. 3, which agrees well with the numerically obtained pattern. The above procedure works independently of the parameter values, demonstrating its validity.

Note that the spatial periodic pattern is obtained as an attractor of the spatial map (\ref{eq:spatialmap}).  Hence, independently of the initial condition in the map, the same periodic pattern is selected as long as the initial condition in the spatial map belongs to the basin of the above attractor. The initial condition in the spatial map corresponds to the value of fixed boundary. Hence, a pattern of the same wavelength is reached independently of the boundary value, while the phase of wave pattern is different, which is consistent with the numerical result (see Fig. 3).

In the state space, pattern selection using the spatial map can be described as follows.
When the diffusion term is small, the shift of the nullcline is small. The same stable fixed point, i.e., $Y_{l}$ on the same branch, continues to exist and is selected according to the above criteria. With iteration of the spatial map, the shift of the nullcline due to the diffusion term is accumulated and at some point, the fixed point of the same branch vanishes with bifurcation, or do not fulfill selection principles.
The selected fixed point is replaced by that of the new branch, as the solution according to the above criteria. Repeating the above procedure, a periodic pattern is obtained. 

This procedure suggests there can be a relationship between the period of oscillation $T$ and the wavelength $\lambda$. 
First, by rescaling the spatial scale, the wavelength $\lambda$ is proportional to $\sqrt{D}$. (The period is scaled by $\tau$ if $f$ and $g$ are similarly scaled by $1/\tau$, but this is already set to unity in (\ref{eq:reaction-diffusion}).)
Now, following the vector field $(f(X,Y),g(X,Y))$ in the state space (Fig. 2A), $(X(t),Y(t))$ oscillates with the period $T$.
On the other hand, when a stripe is formed, the ``spatial orbit'' $(X(x),Y(x))$ along space $x$ orbits the state space similarly to the limit cycle orbit.
Indeed, except for the vicinity of the Hopf bifurcation, we numerically confirmed that one stripe is generated in correspondence with one period of oscillation. 
With the parameter change, the (magnitude of) vector $(f,g)$ in the state space changes, which changes the rate of change along the limit cycle as well as that of the ``spatial orbit'' in the same way.
If these orbits do not have a small radius in the state space, both speeds are expected to change in proportion, implying that the period $T$ and wavelength $\lambda$ (or $\lambda/\sqrt{D}$) change in proportion.
Of course, this is a rough estimate, but this proportionality approximately holds against changes in the parameter values $\beta$ (and $D$), as shown in Fig. 4. 

The fixation of the temporal to the spatially periodic pattern as well as the analysis using the spatial map can be generally applied to a one-dimensional reaction-diffusion equation that has a uniform limit-cycle attractor.  As another illustration, we examined the so-called Brusselator\cite{nicolis1977self}: 
\begin{equation}
\begin{array}{l}
f(X,Y)=BY-XY^2\\[8pt]
g(X,Y)=A-(B+1)X+XY^2\\
\end{array}
\label{eq:brusselator}
\end{equation}

In this case, a uniform periodic oscillation is replaced by a periodic spatial pattern by applying the fixed boundary condition $X(0)=X_0$ for a certain range of parameters, which agrees with the prediction using the corresponding spatial map (see Supplemental Figure 2).

The approximate proportionality between the period and wavelength is worse than the model (2) (see Fig. 4), possibly because the change in the flow $(f(X,Y),g(X,Y)$ against the parameters is highly dependent on the state $(X,Y)$.

In the present letter, we have studied the formation of a periodic pattern from a uniform oscillatory state induced by the boundary conditions. The emergent pattern is predicted as an attractor of the spatial map under the selection principle.
We have confirmed the generality of this pattern formation in reaction-diffusion systems. The periodic oscillation exists in a two-component reaction system with negative feedback. That is, in a system with an activator and an inhibitor, for a certain range of parameters, the spatially periodic pattern that exists as an attractor replaces the uniform oscillatory state, triggered by the fixed boundary condition, if the diffusion of the inhibitor is sufficiently large. 

We also note that the pattern formation process as well as the nullcline analysis is still valid even for $D_Y \neq 0$, although the spatial map explicitly includes the inverse function of $Y$, which would make the analysis more difficult. 
\par Pattern formation generally depends on the boundary and initial conditions. In the spatial map, the former is represented by the initial condition in the map, while the latter is considered by the second and third criteria of the selection principle, which indicate that the initial condition is not far from the uniform oscillatory state.
If the initial condition is far from uniformity, local inhomogeneity can grow. In the state-space representation, whether such growth occurs is determined according to the size of the spatial diffusion term, specifically whether the spatial diffusion term is large enough to go across the nullclines and to induce saddle-node bifurcation. In other words, the condition close to uniformity represents the absence of such inhomogeneity. 
On the other hand, for such initial conditions to allow for the local growth in inhomogeneity, the selection criteria are replaced so that the choice of the fixed-point solution is close to a given initial condition, leading to a requested switch to a different branch of fixed-points. Hence, the choice of the selection principle corresponding to the initial condition can predict the emergent pattern. Control of an emergent pattern is thus possible by manipulating the initial condition.
\par Here, one stripe is formed with one-to-one correspondence of one oscillation period in time. Near the Hopf bifurcation point, however, a complex pattern with one stripe is observed per few periods of oscillation. Indeed, the spatial map can have an attractor with such complex (or quasiperiodic) oscillation in the vicinity of the bifurcation and also by modifying the selection principle. Analysis of such complex patterns in terms of the spatial map (see also \cite{willeboordse1995pattern}) will be of interest in the future.
\par Experimental confirmation of the present pattern formation will be possible in reaction-diffusion systems. In particular, relevance of boundary condition to pattern selection should be of importance because the real experimental system is finite and often under a fixed boundary condition. By carefully examining the boundary effects, we can confirm the present pattern formation mechanism, where the period-wavelength relationship (Fig. 4) will be confirmed.
\par As mentioned in the introduction, spatial pattern formation based on temporal oscillation is often observed in biological morphogenesis\cite{palmeirim1998uncoupling,pourquie2001vertebrate} as well as in the numerical evolution of morphogenesis\cite{fujimoto2008network,kohso2016evo}, where the relevance of cell-cell interaction has been recently discussed in addition to the external morphogen gradient\cite{cotterell2015local,kohso2016evo}. Considering the simplicity in our mechanism, which only requires diffusion of the inhibitor and a fixed boundary, we expect that it could be adopted in biological development, which will be confirmed by examining cell-cell interaction and boundary effects.

The authors would like to thank Nen Saito for useful discussions. This work was partially supported by a Grant-in-Aid for Scientic Research (No. 21120004) on Innovative Areas ``Neural creativity for communication" (No. 4103) and the Platform for Dynamic Approaches to Living Systems from MEXT, Japan.\clearpage
\clearpage

\section*{Figure Legends}
\begin{figure}[htbp]
\begin{center}
\includegraphics[width=15cm]{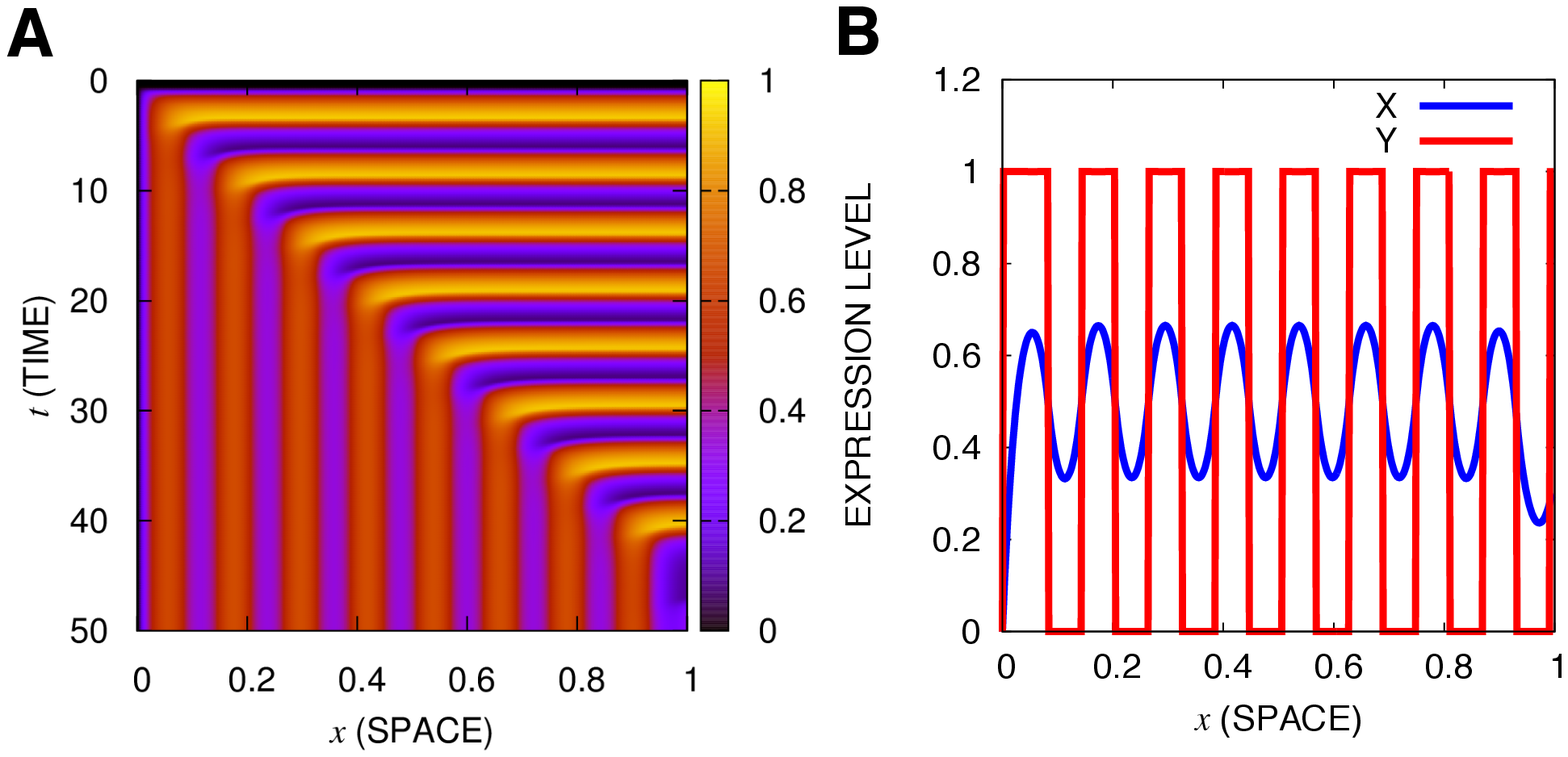}
\caption{Time development of the model (2) under a fixed boundary condition, $X(0)=0$. (\textbf{A}) $X(x,t)$ is displayed with a color scale given by the side bar. The abscissa is time $t$, and the ordinate is space $x$. (\textbf{B}) The generated stationary pattern $X(x)$ and $Y(x)$. The abscissa is the concentration of $X(x)$ or $Y(x)$, and the ordinate is space $x$.}
\end{center}
\end{figure}
\begin{figure}[htbp]
\begin{center}
\includegraphics[width=15cm]{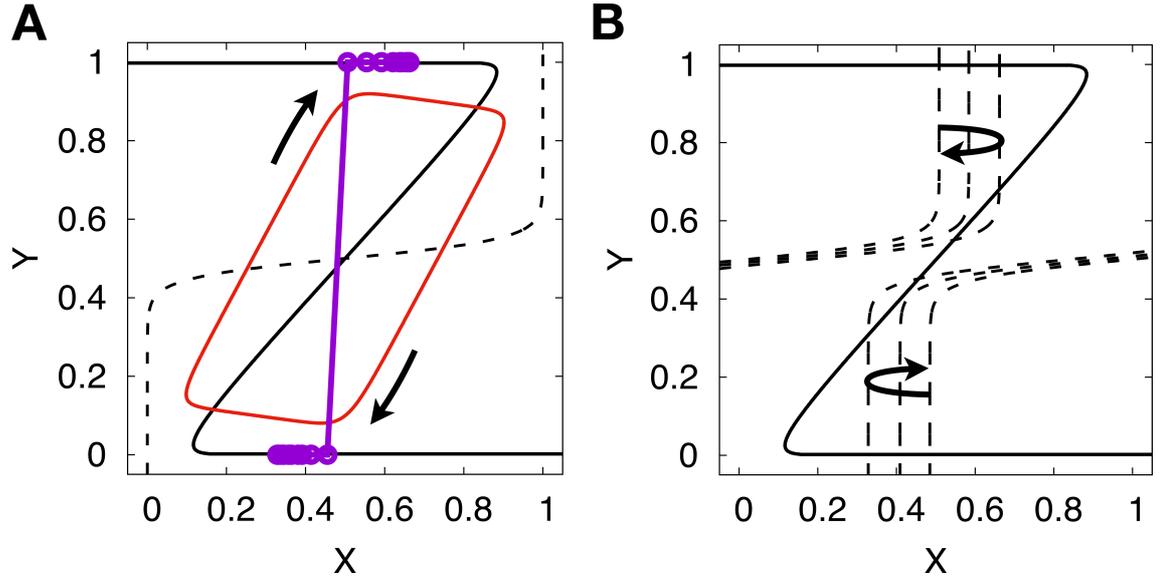}
\caption{
(\textbf{A}) State space plot of nullclines $f(X,Y)=0$ (black solid) and  $g(X,Y)$ (black dotted) as well as the limit cycle orbit (red) are plotted for the model (2) with $D=0$.
(\textbf{B}) State space plot of the nullcline for X ($f +D\partial ^2 X/\partial x^2 =0$) are plotted with the change in space $x$, following the change in the diffusion term $D\partial ^2 X/\partial x^2$, where the cross-points of the nullclines give $(X_{st}(x), Y_{st})(x)$. The nullclines are shifted horizontally with the spatial gradient.
}
\end{center}
\end{figure}
\begin{figure}[htbp]
\begin{center}
\includegraphics[width=15cm]{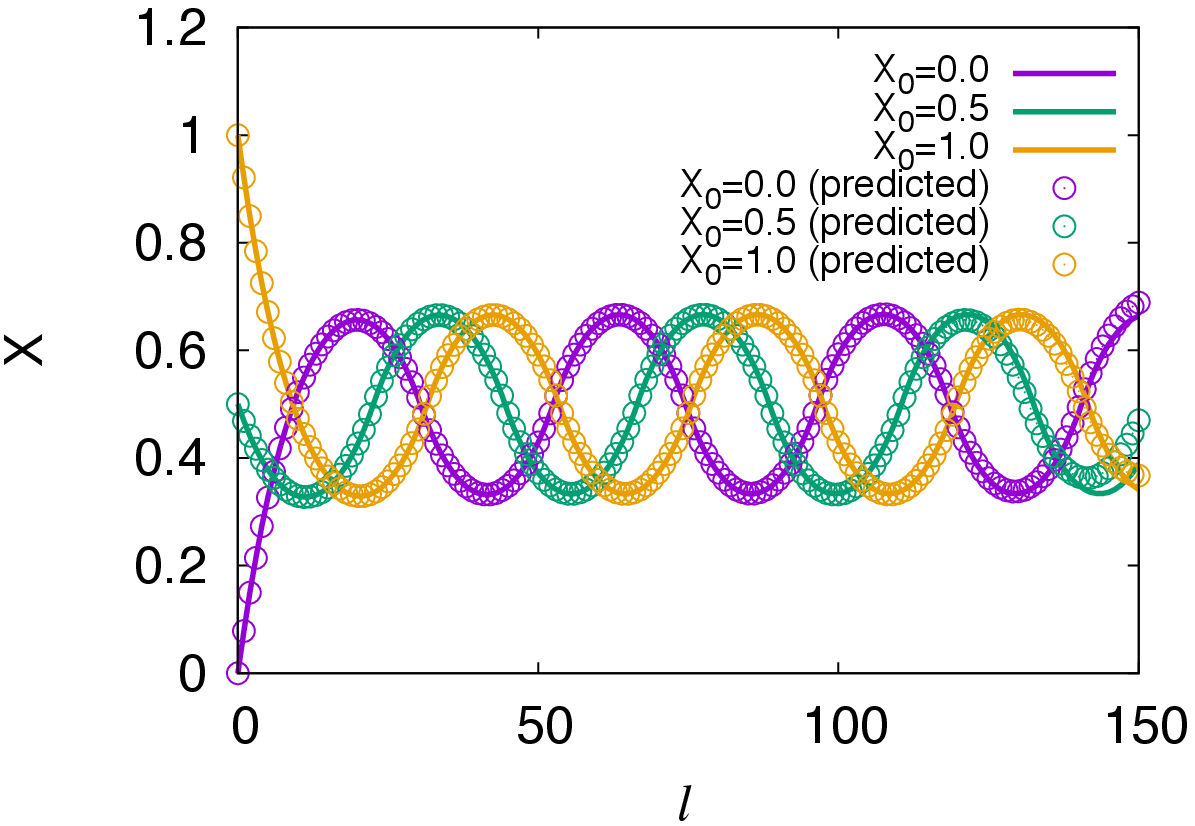}
\caption{Stationary pattern $(X_{st}(l), Y_{st}(l))$ of the reaction diffusion (\ref{eq:example1}) and the predicted pattern of $(X_l, Y_{l})$ obtained from the spatial map (\ref{eq:spatialmap}). The predicted pattern (given by $\circ$) agrees well with the stationary pattern (lines). Three patterns with boundary values of $X_0 = 0.0$ (purple), $X_0 = 0.5$ (green), and $X_0 = 1.0$ (yellow) are considered.
}
\end{center}
\end{figure}
\begin{figure}[!!!!hhhht]
\begin{center}
\includegraphics[width=15cm]{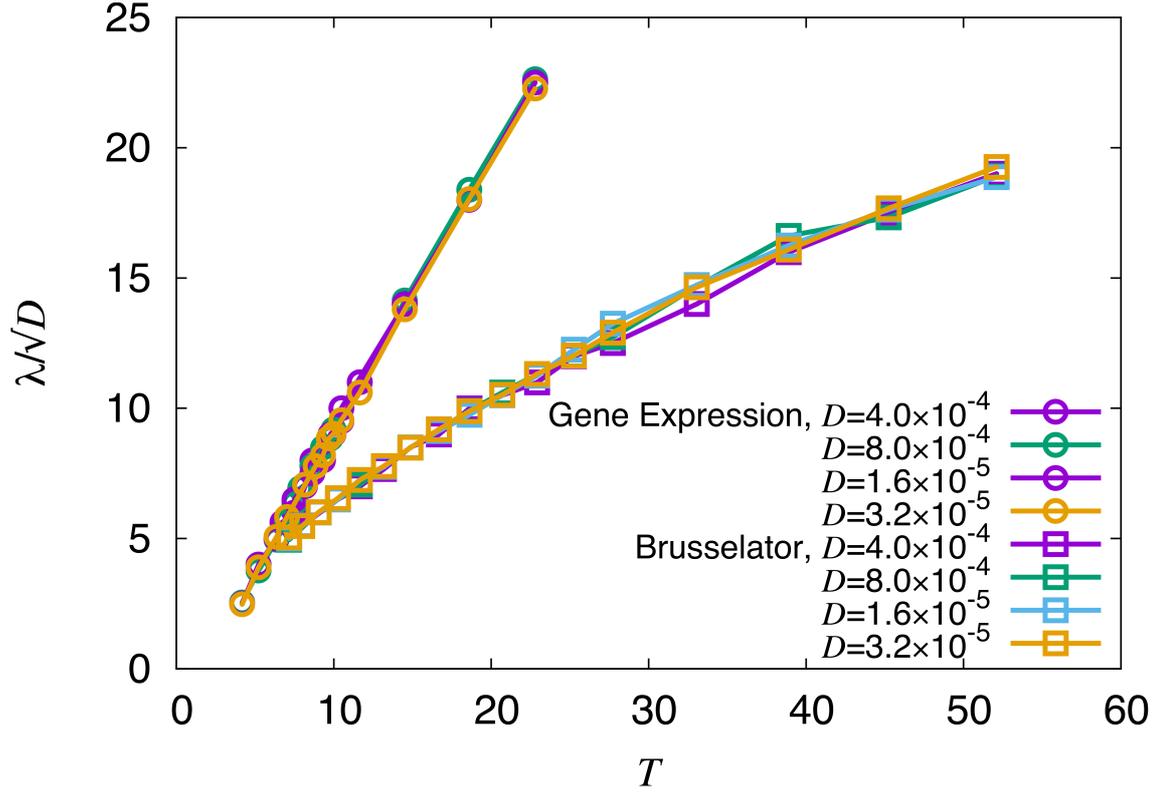}
\caption{Relationship between the scaled wavelength $\lambda/\sqrt{D}$ of the emergent pattern and the period of the limit cycle orbit $T$ is shown by changing parameter $\beta$. Results from different $D$ values are also plotted with different symbols. 
The relationship from the Brusselator Eq. (\ref{eq:brusselator}) is also plotted from the obtained data by varying $B$.}
\end{center}
\end{figure}
\clearpage
\section*{Supplemental Materials}
\begin{figure}[htbp]
\begin{center}
\includegraphics[width=15cm]{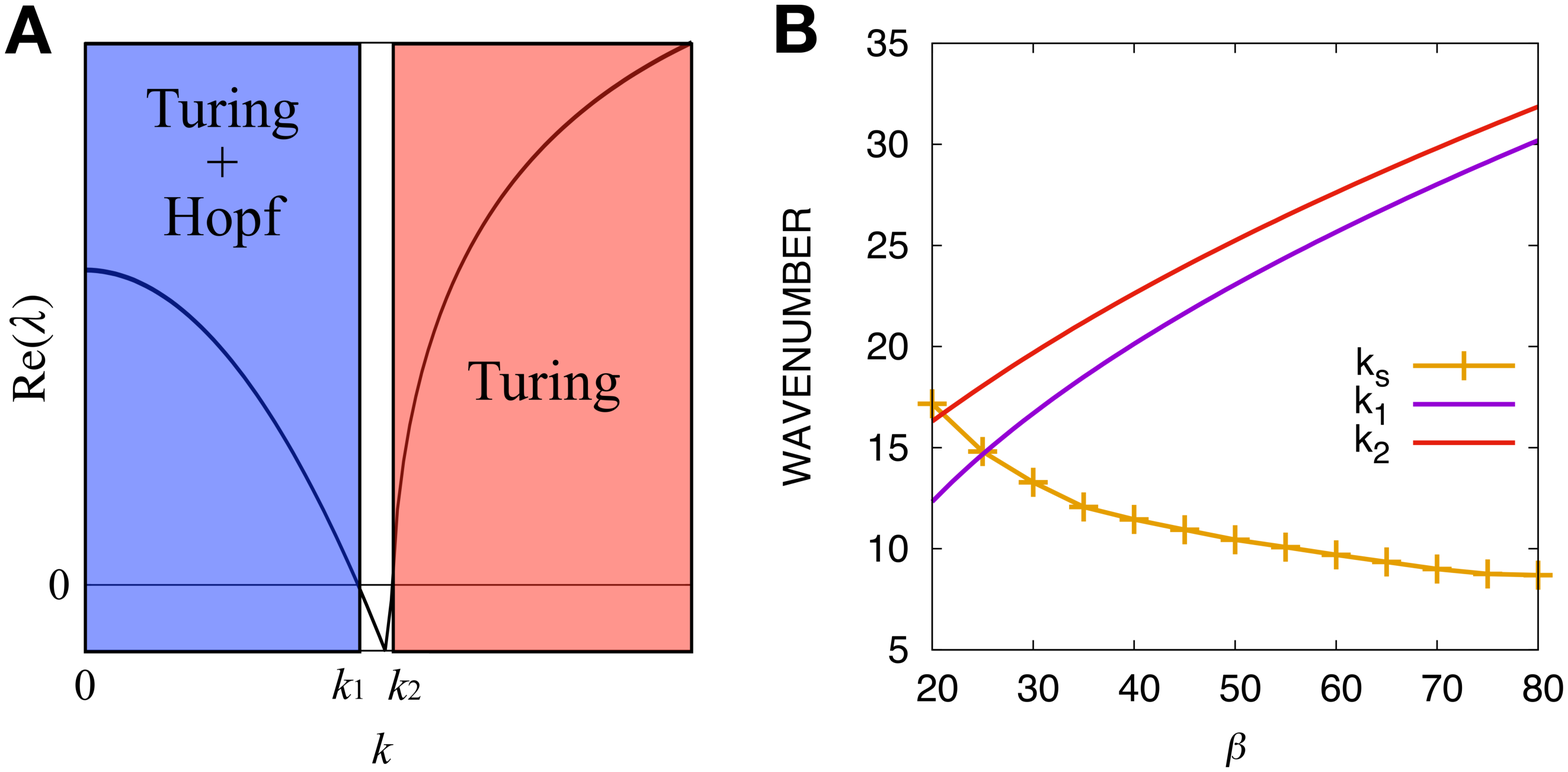}
\begin{quote}
\textbf{SF1} (\textbf{A}) The real part of the Jacobi matrix eigenvalues around the uniform solution for each wavenumber $k$ for model (2), when $\beta > 8$. The eigenvalues are positive for $0< k < k_1$ (where the imaginary part is nonzero) and for $k> k_2$ (where the imaginary part is zero). The imaginary part of the eigenvalue vanishes at $k=\frac{1}{4\pi}\sqrt{\frac{\beta}{D}}<k_2$. In the blue region, both Turing and Hopf instability exist, and only Turing instability exists in the red region. (\textbf{B}) The emergent wavenumber $k_s$ (green, +), marginal wavenumber for the blue region in (\textbf{A}) $k_1$(blue), and marginal wavenumber for the red region in (\textbf{A}) $k_2$ (red) are plotted as a function of $\beta$.
\end{quote}
\end{center}
\end{figure}

\begin{figure}[htbp]
\begin{center}
\includegraphics[width=15cm]{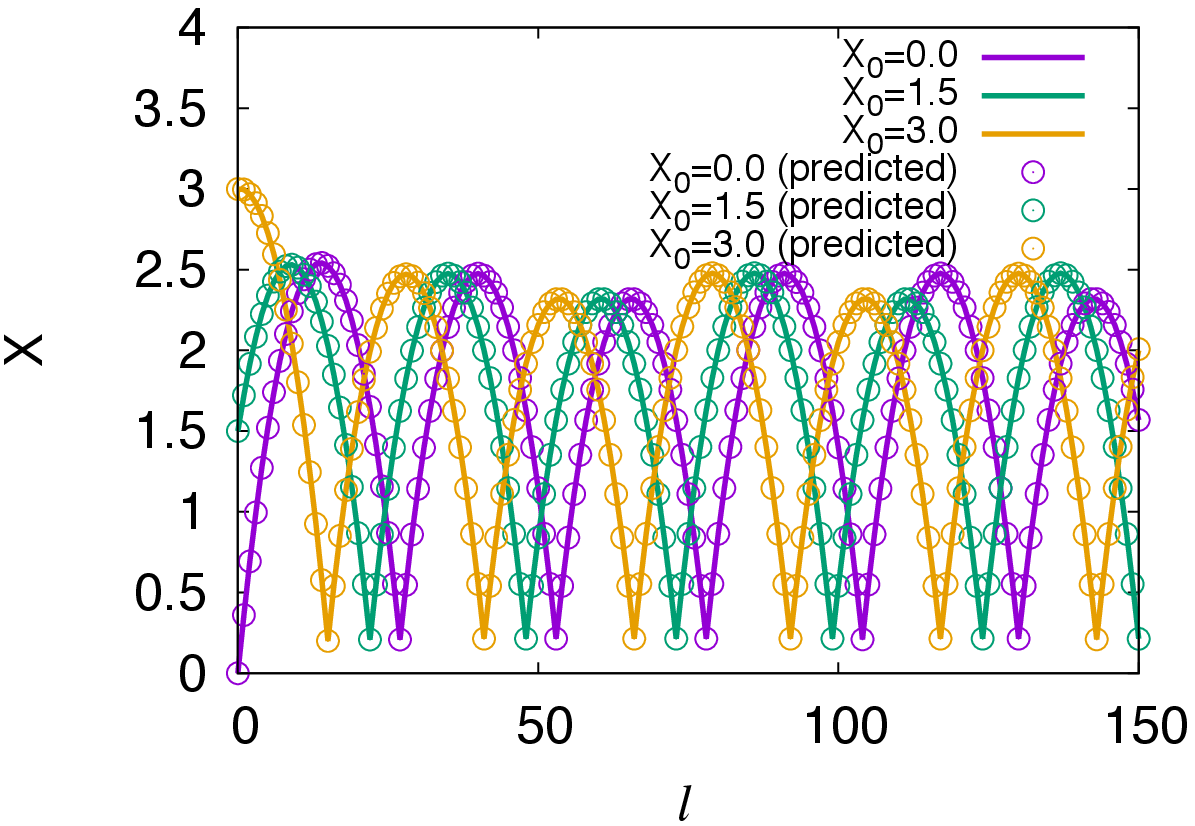}
\begin{quote}
\textbf{SF2} Stationary pattern $(X_{st}(l), Y_{st}(l))$ of the reaction-diffusion (5) and the predicted pattern of $(X_l, Y_l)$ obtained from spatial map (4). The predicted pattern (given by $\circ$) agrees well with the stationary pattern (lines). Three patterns with boundary values of $X_0 = 0.0$ (purple), $X_0 = 1.5$ (green), and $X_0 = 3.0$ (yellow) are considered.

\end{quote}
\end{center}
\end{figure}

\end{document}